\documentstyle[epsfig]{neu}
\begin{document}

\title{%
POLAR CAP MODEL FOR PULSAR HIGH-ENERGY EMISSION}

\author{Alice K. HARDING \& Alexander G. MUSLIMOV$^{\dagger }$\\
{\it Code 661, LHEA, NASA/GSFC Greenbelt, MD 20771 USA,\\
harding@twinkie.gsfc.nasa.gov \& muslimov@lhea1.gsfc.nasa.gov}\\
$^{\dagger}${\it NRC/NAS Senior Research Associate}}

\maketitle

\section*{Abstract}

     The study of physical processes associated with particle
     acceleration in the open field line region above the polar 
     cap (PC) of an isolated neutron star (NS) plays a fundamental 
     role in our understanding and interpretation of high-energy  
     emission from pulsars. The systematic study of particle 
     acceleration and formation of electron-positron pair fronts 
     above the PCs of NSs was initiated two decades ago. The 
     detailed analysis of these processes is now possible with 
     the development of pair cascade codes that enables us to 
     calculate the spectra and pulse profiles of 
     high-energy emission from pulsars. The calculation of pair
     formation and $\gamma$-ray production is being improved to
     include new results on the PC physics. We briefly outline the 
     current status of the PC model for pulsar high-energy emission, 
     focusing on some of our most recent results on the theoretical 
     modeling of the PC acceleration and $\gamma$-ray emission. 

\section{Introduction}

The small but growing subpopulation of $\gamma$-ray active radio pulsars 
raises our hopes in advancing the theoretical models of pulsar 
emission. In this paper we concentrate on the picture in which 
both the particle acceleration and $\gamma$-ray emission occur in
the inner magnetosphere of a neutron star (NS) above the magnetic
pole. The corresponding models, known as polar cap (PC) models, favor
relatively small obliquities (single pole emission), and they
crucially depend on curvature radiation (CR) and on the QED process of
magnetic pair creation. The prototype of the PC models has been suggested by 
Ruderman \& Sutherland (1975). Some outstanding problems and progress 
in this field were addressed clearly by Curtis Michel, Jonathan 
Arons, and other speakers at the conference. In this written report we  
will schematically describe the relevant results of our most 
recent work that are being prepared for publication in a more complete
form. The bottom line of this presentation is that 
{\bf the PC model for pulsar high-energy emission is generally a
viable approach, but it needs to be revised and improved in a number
of important theoretical aspects.}

\section{Observation Summary}

Let us briefly summarize the main observational facts that need to be 
explained by any model of $\gamma$-ray emission from pulsars (see 
Thompson et al. 1997 for a review). First, most $\gamma$-ray
light curves show two peaks (separated by 0.4-0.5 in phase) with 
bridge emission. These peaks are in phase with the similar hard X-ray 
peaks. Second, in the high-energy band (few to tens GeV) of the
spectra of $\gamma$-ray pulsars there seem to be spectral breaks or
cutoffs (the exception is PSR 1509-58 for which this occurs between 4
and 30 MeV). Third, there is a systematic variation
of spectral hardness through the $\gamma$-ray pulse profile. In 
Geminga this hardness variation seems to be more complex or even
behave in the opposite sense than in other pulsars for which such a 
study has been performed. And fourth, albeit more ambiguous,  
is that the hardness of the phase-averaged spectra tend to increase 
with the pulsar spin-down age. In addition, for some $\gamma$-ray
pulsars (e.g. Geminga, Vela, and PSR 0656+14) the soft component of
the X-ray spectrum has a broad profile centered between the
$\gamma$-ray peaks. Finally, in all $\gamma$-ray pulsars excluding
Crab, there is a phase offset between the corresponding peaks at 
different wavelengths, from radio to gamma.

\section{Basics of the PC Model and Main Input Physics}

Several types of pulsar models have studied particle
(electron/positron) acceleration due to charge deficits at different 
locations in the neutron star magnetosphere. Polar cap (PC) models 
consider the formation of a parallel electric field, $E_{\parallel } 
\equiv {\bf E}\cdot {\bf B}/B$, in the open field region near the 
magnetic poles, while outer gap
models consider accceleration in the outer magnetosphere, near the
null charge surface (see Mestel 1998 for the most recent and
comprehensive review of pulsar electrodynamics).  

The initial calculations of electron-positron pair formation fronts 
(PFFs) assumed that the primary electrons began accelerating at the 
NS surface and that curvature radiation (CR) was the only mechanism for 
providing pair-producing photons (Arons 1983). In recent years, it has 
become clear that inverse-Compton scattering (ICS) of soft thermal X-ray 
photons from the hot NS surface by the primary electrons is also an 
important mechanism above the PC. As well as being a
significant energy loss (Xia et al. 1985, Daugherty \& Harding 1989, 
Sturner 1995) and radiation (Sturner \& Dermer 1994) mechanism,
ICS can also provide photons capable of producing
pairs. The standard models of PC acceleration thus need substantial 
revision.

We use the electric field due to inertial frame dragging above the
NS surface, first calculated by Muslimov \& Tsygan (1990, 1992;
hereafter MT90, 92). The regime under which the generation of this 
field occurs is actually the same as implied in the electric field 
computations by Arons \& Scharlemann (1979; hereafter AS79), assuming 
space-charge limited flow in flat space.  An electric field must be 
present above the NS surface because as charges flow along open
magnetic field lines, the corotation, or Goldreich-Julian charge 
density $\rho_{\rm GJ}$, cannot be maintained. Even though the actual 
charge density $\rho = \rho_{\rm GJ}$ and therefore $E_{\parallel} =
0$ at the surface, the curvature of the field lines causes the
area of the open field region, through which the particles flow, to
increase faster with height than $\rho_{\rm GJ}$, and a charge deficit 
grows.  Thus, the $E_{\parallel}$ cannot be shorted-out and grows
with altitude up to about one stellar radius above the surface, and then
it gradually declines. General relativity causes a significant
modification (MT90, 92) of this induced electric field,
through the effect of dragging of inertial frames, a consequence of
the distortion of spacetime by a rotating gravitating body. Thus the
Goldreich-Julian charge density, which is the charge density required the
make magnetospheric particles drift in corotation with the star, will
differ from that in flat space.  This charge difference enhances 
$E_{\parallel}$ over what it would be in flat space, by a factor of 50
- 100 for a typical 1 s pulsar. The frame-dragging contribution (see
e.g. first term in Eqn [1]) to $E_{\parallel}$ depends on $\cos\chi$, 
where $\chi$ is the pulsar obliquity angle.  Particle 
acceleration may therefore occur throughout the entire open field line 
region, with the relative contribution of the frame dragging component
to $E_{\parallel}$ being strongest for pulsars with small
obliquities. In our modeling we 
properly incorporate the effect of the screening of $E_{\parallel }$
at some height above the stellar surface by creation of
electron-positron pairs in the strong magnetic field. 
For illustration we now present an explicit formula for the case where 
the screening of $E_{\parallel }$ occurs at altitude of order of the 
PC size,
\begin{equation}
E_{\parallel } \simeq - {3\over 2} \left( {{\Omega R}\over c} \right)
{{B_0}\over {1-\varepsilon}} \left( 1-{z\over z_0} \right) z 
\left[\kappa \cos \chi + {1\over 2} \theta _0 \xi H(1) \delta (1)
\sin \chi \cos \phi \right] ,
\end{equation}
where $\Omega $ is the pulsar rotation frequency, $B_0$ is the surface
value of the magnetic field strength at the magnetic pole,
$\varepsilon = r_{\rm g}/R$, $r_{\rm g}$ is the gravitational radius
of the NS, $z$ is the altitude scaled by the stellar radius
$R$, $z_0$ is the altitude at which the electric field is fully
screened out, $\kappa = \varepsilon I/MR^2$, and $I$, and $M$ are the
moment of inertia and mass of the NS, respectively, 
$\xi \equiv \theta /\theta _0$ is the magnetic colatitude scaled by 
the half-opening angle of the PC, $\phi $ is the azimuthal
angle, and the functions $H$ and $\delta$ are the correction factors
due to the strong gravity, with $H(1)\delta(1) \simeq 1$.

\section{Modeling of the PFFs and  $\gamma$-ray Emission}

Another effect which has never been included in PC acceleration 
models is the formation of a lower PFF due to the positrons that are
turned around and accelerated downward from the electron PFF.
Although the number of positrons which are accelerated downward is
small compared to the primary current and even to the charge deficit at the
upper PFF, the multiplicity of the downward cascades is quite large. 
Thus, the amount of charge
produced by only a small number of downward moving positrons may be 
sufficient to establish a second PFF.  Although downward going 
cascades have been discussed in previous papers (AS79), their effect 
on the acceleration of primaries has not been investigated.

In our most recent study (Harding \& Muslimov 1998b; hereafter HM98b) 
we investigate the acceleration of primary electrons and secondary 
(downward-moving) positrons above a pulsar PC, assuming space-charge 
limited flow (free emission) of particles from the NS surface. Both 
electrons and positrons suffer energy loss and emit photons from CR
and ICS. 
\begin{figure}[t] 
\vskip -1.5cm
\epsfig{file=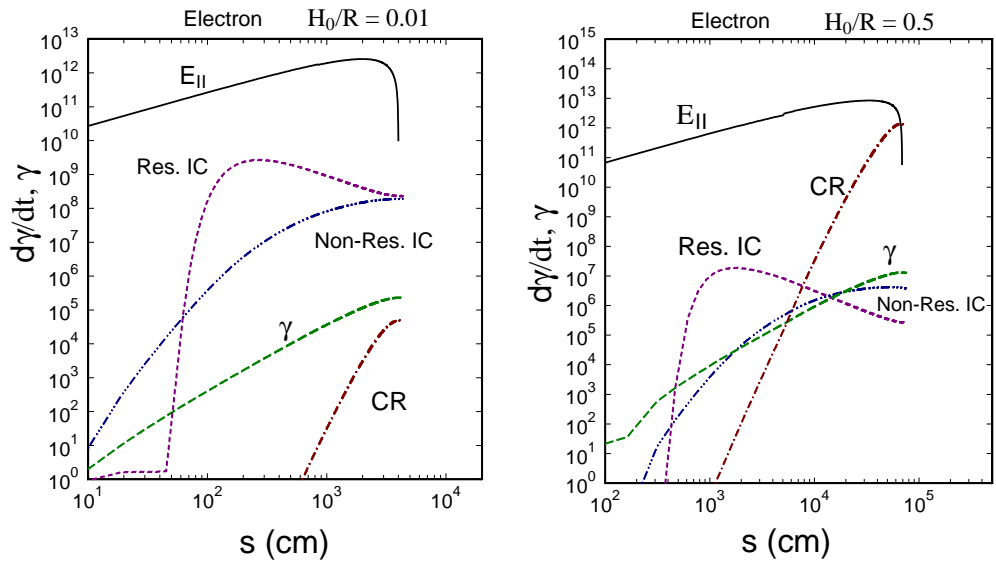,height=3.25in}
\vskip -0.7cm
\caption{Left: $E_{\parallel }$, Lorentz factor, $\gamma$, 
and energy loss-rates ($\dot{\gamma }$) due to non-resonant/resonant ICS 
and CR as functions of the acceleration lengths. Right: 
The same as in Left, but for the CR-controlled PFFs. Here $H_0$ is the
height of lower PFF (start of acceleration) in units of $R$.}
\label{fig1}
\end{figure}
\noindent
In Fig. 1 we summarize some of our calculations for the electrons to 
illustrate the self-consistent solutions of the PFFs 
controlled by ICS and CR. We compute the location of 
both electron and positron PFFs due to one-photon pair production as a 
function of magnetic colatitude and height of the lower PFF. When the 
electrons are assumed to accelerate from the NS surface, we find that 
ICS photons produce the PFFs, in agreement with the results of 
Zhang \& Qiao (1996). However, we also find that there is substantial 
asymmetry between the scattering of upward-going electrons and 
downward-going positrons by the same thermal X-ray photons: the
electrons scatter these photons at angles less than $\pi/2$, while the 
positrons scatter the photons head-on.  The photons scattered by
positrons of the same energy are therefore more energetic and produce 
pairs in a shorter distance.  These pairs may poison $E_{\parallel }$  
up to some altitude above the surface.  We demonstrate that (see 
Fig. 1) the double PFFs controlled by CR tend to 
set up at much higher altitudes than those controlled by ICS. This 
fact is essential for reproducing the widely 
separated double-peaked $\gamma$-ray (and also hard X-ray) profiles 
observed in 
$\gamma$-ray pulsars. Also, we suggest that stable, double PFFs can 
form only when CR photons produce them; i.e. at 
a height where electron CR losses overtake ICS losses (see Fig. 1, Right
panel). One interesting result of our computations is that
the accelerated voltage limited by CR-controlled PFFs is only a 
function of magnetic colatitude 
(i.e. geometry of the open field lines), ranging between $\sim 10^7$
and $3\cdot 10^7\, mc^2$, and is insensitive to pulsar parameters such as
period and surface magnetic field and even to the height of the
acceleration. 
\begin{figure}[t] 
\vskip -1.5cm
\centerline{\epsfig{file=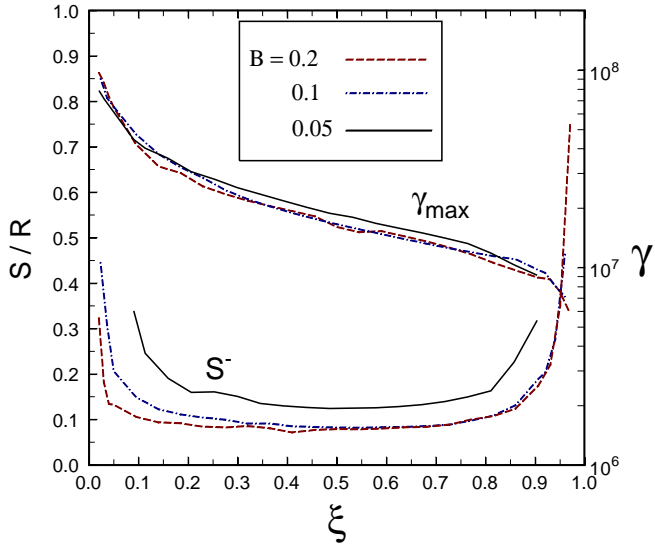,height=3.5in}}
\vskip -0.7cm
\caption{Profiles of the maximum electron Lorentz factor and 
the total acceleration length of the CR-produced upper pair front 
(see Fig. 1, Right panel) across the PC. The magnetic field strength 
is in units of $B_{\rm cr}\simeq 4.4 \cdot 10^{13}$ G.}
\label{fig2}
\end{figure}
\noindent
In Fig. 2 we present the profiles of the electron upper 
PFF controlled by CR (i.e. in the regime of electron acceleration
where the CR losses dominate the ICS losses) and the corresponding
maximum Lorentz factor as functions of coordinate $\xi $ (see the
notations following Eqn [1]). Figure 2 illustrates that the 
electron maximum Lorentz factor (total voltage drop) decreases toward 
the PC rim, $\xi = 1$, because of decrease in $E_{\parallel}$, while
the height of the PFF increases at the magnetic axis, because of 
significant increase in the pair-production attenuation length. 
The stable location of the lower PFF depends primarily
of surface magnetic field, and PC size and temperature, ranging between 
0.5 and 1.0 stellar radius, but is insensitive to period.

\section{On the PC Heating and Surface X-ray Emission}

Some fraction of positrons produced e.g. in the first cascades may
return toward the PC surface (an accurate determination of the
fraction of backflowing positrons requires a self-consistent
calculation of the screening of $E_{\parallel }$ by the cascade pairs
and the modification of the pair spatial distribution by the screened 
$E_{\parallel }$). These backflowing positrons slightly suppress the voltage 
all the way down to the bottom of the polar magnetic flux tube, 
so that the electric field $E_{\parallel }$ (and also potential $\Phi
$) vanishes at the height $H_0$ above the stellar surface rather than
at the actual stellar surface. This occurs because the flux of 
backflowing positrons is equivalent to the corresponding enhancement 
of the total electron current and therefore of the maximum number of 
electrons per second to be ejected into the acceleration region. 
Thus, to satisfy the zero-electric field boundary condition the 
ejection radius should fix itself at the larger value corresponding 
to the higher rate of supply of Goldreich-Julian charge. It is 
interesting that, just by using very general reasoning, we can derive 
an upper limit on the fraction of returning positrons. 

The maximum possible total power put into the backflowing positrons can be
estimated as (cf. Muslimov \& Harding 1997; hereafter MH97, Eqns
[76]-[78]; and see HM98b):
\begin{equation}
\left\{ L_{\rm e^+} \right\} _{\rm max } \approx \lambda ~L_{\rm sd},
\end{equation}
where 
\begin{equation}
\lambda \simeq (3/4) x^3 \left( 
\kappa ^2 x^3 \cos ^2 \chi + (3/16) H^2 \theta^2 \sin^2 \chi \right),
\end{equation}
and $L_{\rm sd}$ is the spin-down luminosity of a pulsar,
$x=1/(1+H_0/R)$, $H_0$ is the altitude of the lower (positron) PFF 
(see Fig. 1). 
For relatively small obliquities and typical pulsar
spin periods of 0.1 - 1 s the first term in Eqn (3) dominates,
and we get 
\begin{equation}
\left\{ L_{\rm e^+} \right\} _{\rm max } \approx 
(3/4) \kappa ^2 x^6 L_{\rm sd}.
\end{equation}
For the 1.4 $M_{\odot }$ NS and for a broad range of realistic
equations of state of dense matter (see e.g. Ravenhall \& Pethick 
1994 for the calculations of the
NS moment of inertia for various equations of state) $I/(MR^2) \approx 
(0.2-0.25) (1-r_{\rm g}/R)^{-1}$. Thus, for the NS of 1.4 $M_{\odot }$ 
and 8-10 km radius (which is consistent with the realistic stellar
models) we can estimate 
$\kappa \approx 0.15-0.27$. Given $x\approx 0.5-1$, Eqn (4) yields
$\lambda \approx (3\cdot 10^{-4}-2\cdot 
10^{-2} ) (\kappa/0.15)^2$. This estimate combined with 
the estimated X-ray luminosities (that imply isotropic emission) of
pulsars (see Becker \& Tr\"{u}mper 1997) may have rather interesting 
implications. 

For example, if the X-ray fluxes in some of 
these pulsars are dominated by the photons from the heated  
(e.g. by the backflowing positrons) PC, then this may  
indicate that these pulsars operate in an oscillatory regime 
(see Sturrock 1971, and also HM98b). Alternatively, we should note 
that according to the estimates by Becker \& Tr\"{u}mper the 
pulsed X-ray luminosities (for the case of isotropic emission), 
$L_{\rm x}^{\rm p}$, for e.g. Crab, Vela, Geminga, and PSR 0656+14 are, 
respectively, $1.6\cdot 10^{-3}$, $0.7 \cdot 10^{-5}$, 
$1.3\cdot 10^{-4}$, and $3.7\cdot 10^{-3}~L_{\rm sd}$. 
If we assume that at least for Vela, Geminga, and
PSR 0656+14 (see e.g. HM98a for the modeling of
the soft X-ray and $\gamma-$ray emission for these pulsars) the X-ray 
emission is beamed into a solid angle of $\sim $ 1 steradian, then for
these pulsars we can estimate that $\lambda _{\rm x_{\rm
anis}}^{\rm p} \equiv L_{\rm x_{\rm anis}}^{\rm p}
/L_{\rm sd} \rm \sim 6 \cdot 10^{-7}$, $10^{-5}$, and 
$3\cdot 10^{-4}$, respectively. The corresponding PC temperature for 
these pulsars can be estimated as $T_{\rm pc} \sim (0.6-1) \cdot 10^6 
(10^5 \lambda _{\rm x_{\rm anis}}^{\rm p})^{1/4}(B_0/
4\cdot 10^{12}~{\rm G})^{1/2}(R/8~{\rm km})^{3/4}(P/0.1~{\rm s})^{-3/4} 
~{\rm K}$. Thus, the estimated value of $\lambda _{\rm x_{\rm
anis}}^{\rm p} \sim (3\cdot 10^{-5}-10^{-2})~\lambda $ may indicate
that the backflowing
positrons precipitate onto the effective area smaller than that of the
standard PC (e.g. the returning positrons focus toward the 
magnetic axis, HM98b), and/or that the fraction of the
returning positrons is well below the maximum possible one. Then the 
latter would support the quasi-steady state (MH97) rather than the 
oscillatory (Sturrock 1971) regime of pulsar operation.   

\section{Conclusions}

In conclusion we emphasize that the PC model is capable of
explaining and reproducing (see Harding 1996, Daugherty \& Harding
1996) the main observational facts mentioned in $\S~2$: the widely separated 
double-peaked profiles (also in hard X-rays) with bridge component, 
very steep high-energy spectral cutoffs (due to magnetic pair 
production), and systematic soft-hard-soft hardness variation in
the pulse (due to the softening of emerging spectra by cascades). 

The novel developments we would like to introduce here are: 
1) the double PFF controlled by CR, a 
self-limiting buildup where the electrons are accelerated from 
the lower front and a small fraction of positrons returning from the 
upper front to produce the lower front;  2) the establishment of the 
PFFs and pair cascades at higher altitudes; and 3)
the revised treatment of a feedback between the inflow of cascade
particles into the acceleration region and accelerating electric 
field, including the PC heating by returning positrons. 

Although the PC model has had success in accounting for a number of main
observational features of $\gamma$-ray pulsars, there remain some 
puzzling observational data and exceptions (e.g., Crab: alignment of 
$\gamma$-ray pulses with those at other wavebands; Geminga: complex 
behaviour of spectral hardness through the pulse; Geminga, Vela, and 
PSR 0656+14: soft pulsed X-ray emission; etc.). Given the advancement 
in understanding of PC acceleration discussed above, we will be able
to model characteristics of pulsar high-energy emission from PC pair 
cascades in more detail and ultimately address these unsolved problems. 

\section{References}


\re
   Arons, J. 1983, Ap.J., 266, 215
\re
   Arons, J., \& Scharlemann, E.~T. 1979, Ap.J., 231, 854: AS79
\re
   Becker, W., \& Tr\"{u}mper, J. 1997, A\&A, 326, 682
\re 
   Daugherty, J.~K., \& Harding, A.~K. 1989, Ap.J., 336, 861
\re 
   -----. 1996, Ap.J., 458, 278
\re
   Harding, A.~K. 1996, in Pulsars: Problems \& Progress, 
eds S. Johnston, M.~A. Walker, and M. Bailes, ASP Conf. Series, 
V. 105, p. 315 
\re
   Harding, A.~K., \& Muslimov, A.~G. 1998a, Ap.J., in press: HM98a  
\re
  -----. 1998b, Ap.J., submitted: HM98b  
\re
   Mestel, L. 1998, Stellar Magnetism (Oxford: Oxford University Press)
\re
   Muslimov, A.~G., \& Harding, A.~K. 1997, Ap.J., 485, 735: MH97
\re
   Muslimov, A.~G., \& Tsygan, A.~I. 1990, AZh, 67, 263: MT90
\re
   -----. 1992, MNRAS, 255, 61: MT92
\re
   Ravenhall, D.~G., \& Pethick, C.~J. 1994, Ap.J., 424, 846
\re
   Ruderman, M.~A., \& Sutherland, P.~G. 1975, Ap.J., 196, 51
\re
   Sturner, S.~J. 1995, Ap.J., 446, 292
\re
   Sturner, S.~J., \& Dermer, C. 1994, Ap.J., 420, L79
\re
   Sturrock, P.~A. 1971, Ap.J., 164, 529
\re 
   Thompson, D., et al. 1997, In 4th Compton Symp., ed. C.~D. Dermer,
   M.~S. Strickman, \& J.~D. Kurfess (New York: AIP), p. 39
\re
   Xia, X.~Y. et al. 1985, A\&A, 152, 93
\re
   Zhang, B., \& Qiao, G.~J. 1996, A\&A, 310, 135

\end{document}